\documentclass[journal,onecolumn]{IEEEtran}
\IEEEoverridecommandlockouts
\usepackage{cite}
\usepackage{flushend}
\usepackage{amsmath,amssymb,amsfonts}
\usepackage{algorithmic}
\usepackage{graphicx}
\usepackage{textcomp}
\usepackage{xcolor}
\usepackage{pgfplots}
\usepackage{csquotes}
\usepackage{hyperref}
\usetikzlibrary{mindmap}
\def\BibTeX{{\rm B\kern-.05em{\sc i\kern-.025em b}\kern-.08em
    T\kern-.1667em\lower.7ex\hbox{E}\kern-.125emX}}

\pgfplotsset{compat=1.17}
\usepackage{geometry}
\begin{document}

\title{
Another Systematic Review? A Critical Analysis of Systematic Literature Reviews on Agile Effort and Cost Estimation
\thanks{***This is the author's version of the manuscript accepted for publication in the Proceedings of 2025 ACM/IEEE International Symposium on Empirical Software Engineering and Measurement (ESEM). The copyright owner’s version can be accessed on this \href{https://doi.ieeecomputersociety.org/10.1109/ESEM64174.2025.00051}{link}. This manuscript version is made available under the CC-BY-NC-ND 4.0 license. http://creativecommons.org/ licenses/by-nc-nd/4.0. }
}

\author{\IEEEauthorblockN{Henry Edison and Nauman bin Ali} \\
\IEEEauthorblockA{\textit{Department of Software Engineering} \\
\textit{Blekinge Institute of Technology,} 
Sweden\\
\{henry.edison, nauman.ali\}@bth.se }
}

   

\maketitle

\begin{abstract}
Background: Systematic literature reviews (SLRs) have become prevalent in software engineering research. Several researchers may conduct SLRs on similar topics without a prospective register for SLR protocols. However, even ignoring these unavoidable duplications of effort in the simultaneous conduct of SLRs, the proliferation of overlapping and often repetitive SLRs indicates that researchers are not extensively checking for existing SLRs on a topic. Given how effort-intensive it is to design, conduct, and report an SLR, the situation is less than ideal for software engineering research.
Aim: To understand how authors justify additional SLRs on a topic.
Method: To illustrate the issue and develop suggestions for improvement to address this issue, we have intentionally picked a sufficiently narrow but well-researched topic, i.e., effort estimation in Agile software development. We identify common justification patterns through a qualitative content analysis of 18 published SLRs. We further consider the citation data, publication years, publication venues, and the quality of the SLRs when interpreting the results.
Results:  The common justification patterns include authors claiming gaps in coverage, methodological limitations in prior studies, temporal obsolescence of previous SLRs, or rapid technological and methodological advancements necessitating updated syntheses. 
Conclusion:
Our in-depth analysis of SLRs on a fairly narrow topic provides insights into SLRs in software engineering in general. By emphasizing the need for identifying existing SLRs and for justifying the undertaking of further SLRs, both in design and review guidelines and as a policy of conferences and journals, we can reduce the likelihood of duplication of effort and increase the rate of progress in the field. 
\end{abstract}


\section{Introduction}
Systematic literature reviews (SLRs) were adopted from medicine into software engineering as a tool to establish evidence-based software engineering \cite{kitchenham2015evidence}. The purpose of SLRs is to identify, assess, and synthesize evidence related to narrow and specific research questions reported in empirical studies \cite{kitchenham2007guidelines}. Since its introduction in the software engineering domain in 2004 \cite{kitchenham2009systematic}, the number of SLRs published in top-tier conferences and journals has increased rapidly \cite{ali2018reliability,wang23}.  Several studies have evaluated the methodological rigour and relevance of published SLRs in the field. Such studies have resulted in revised and improved SLR guidelines for designing and conducting \cite{petersen2015guidelines,kitchenham2015evidence}, reporting \cite{ali2018reliability,kitchenham2022segress} and evaluating \cite{AliU19,usman2023quality,ralph2020empirical}.

However, when it comes to the quality assessment of SLRs in software engineering, mainly the DARE criteria\footnote{\href{https://www.crd.york.ac.uk/CRDWeb/AboutPage.asp}{DARE criteria used by NIHR Centre for Reviews and Dissemination}} are used. Costal et al. \cite{costal2021tertiary} found that of the 35 studies they evaluated, only one study used an adaptation of the Assessing the Methodological Quality of Systematic (AMSTAR) framework, and four proposed their own assessment criteria. The overwhelming majority (that is, 30 of the 35 studies) used DARE or its variations. DARE criteria and its adaptations in previous studies have very rightly focused on assessing the methodological rigor and the relevance of SLRs. However, DARE has not questioned whether these SLRs were warranted in the first place or not.

Like all research, SLRs must be carried out diligently, following rigorous methodological guidelines, with results accurately reported and interpreted. SLRs typically require much greater effort than conventional literature reviews. Therefore, researchers in software engineering should carefully consider whether undertaking a systematic literature review is warranted before proceeding. One may assume that it is a given and would be handled by the authors or during the peer review process. We agree with this assumption but would like to empirically assess if that is the case in practice. Some of this unintended redundancy may be unavoidable, due to various reasons, lack of visibility for certain venues, simultaneous efforts undertaking literature reviews with similar or overlapping aims, but we want to ascertain the extent of the overlap and duplication of effort and suggest guidelines to avoid it in the future. 

For this purpose, we have consciously chosen SLRs in the area of effort estimation in Agile, as an illustrative example case. Effort estimation has always been a topic of interest for secondary studies. For example, in one of the earliest tertiary studies identifying early adopters of SLRs in the software engineering community, Kitchenham et al. \cite{kitchenham2009systematic} identified that among the first 20 SLRs, seven were about cost estimation. Since the early 2000s, Agile has had a major impact on software development and on effort estimation. Given the extensive historical emphasis on this area and the continued interest, we decided to use this as a case in this study.

In this study, we pose and answer the following research questions:
\begin{itemize}
\item [RQ1:] Which systematic literature reviews are published on cost/effort estimation in the agile software development context?
\item [RQ2:] What are the characteristics of these SLRs? We focus on salient features of these studies, including aims and objectives, years covered in search, quality scores, e.g., using DARE, publication venue ranking, etc. 
\item [RQ3:] Are the SLRs aware of previous reviews on the topic? We will use citations or discussions of existing reviews to indicate awareness.
\item [RQ4:] What justifications are presented in the SLRs to justify undertaking another SLR on the topic?
\end{itemize}

The remainder of the paper is structured as follows. Section \ref{sec:methodology} describes the data collection process. Then, in Section \ref{sec:results_analysis}, we report the results and analysis, which are discussed in Section \ref{sec:discussion} and concluded in Section \ref{sec:conclusion}.


\section{Research Methodology}
\label{sec:methodology}
We followed Kitchenham et al.'s \cite{kitchenham2015evidence} guidelines to build the sample SLRs and analyze them to answer the research questions. The procedures we followed are detailed in this section.

\subsection{Search strategy}

We used the following search string to identify related SLRs in four databases (see Table \ref{tab:search_results}):
\begin{displayquote}
    \textit{(``systematic review'' OR ``research review'' OR ``research synthesis'' OR ``research integration'' OR ``systematic overview'' OR ``systematic research synthesis'' OR ``integrative research review'' OR ``integrative review'' OR ``systematic literature review'' OR ``literature review'') AND (estimat* OR predict* OR forecast* OR calculat* OR assessment OR measur* OR sizing) AND (effort OR resource OR cost OR size OR metric OR user story OR velocity) AND ( Agile OR ``extreme programming'' OR ``Scrum'' OR ``feature driven development'' OR ``dynamic systems development method'' OR ``crystal software development'' OR ``crystal methodology'' OR ``adaptive software development'' OR ``lean software development'')}
\end{displayquote}

One could even argue that a more inclusive approach would have been not to have the restriction of the third set of keywords capturing the Agile context. Previous reviews on effort and cost estimation may have included literature from this context without making it the central theme, and thus, they may not have mentioned it in the title, abstract, or keywords of a paper. However, after skimming through the titles of the hits, we could see that the search has already been given a sizeable sample of relevant secondary studies. Thus, we decided to keep the rather narrow focus of the search string.

To complement the search, we also searched on Google Scholar using a simple search string: \textit{(``effort estimation'' OR ``cost estimation'') AND agile AND (``systematic literature review'' OR ``systematic review''}. We reviewed the first 50 hits ranked by Google Scholar on their relevance to the search string and identified nine additional SLRs. We believe this additional step adds to the confidence that we have good coverage of SLRs published on cost/effort estimation in agile software development.

We ran the search string on March 11, 2025, and retrieved a total of 162 papers, as shown in Table \ref{tab:search_results}.

\begin{table}[]
    \centering
    \caption{Search results}
    \label{tab:search_results}
    \begin{tabular}{p{5.5cm}p{1cm}p{1cm}p{1cm}}
    \\ \hline
    Database & Retrieved & Duplicates & Remaining \\ \hline
    Scopus & 23 & 10 &  13 \\
    IEEE Xplore & 21 & 1 & 20 \\
    ACM Digital Library & 14 &  & 14 \\   
    ISI Web of Science & 104 & 7 & 97 \\ \hline
    Total & 162 & 18 & 144 \\ \hline
\end{tabular}
\end{table}

\subsection{Selection criteria} 
For a paper to be included, it must meet all of the following:
\begin{itemize}
    \item It reviews the literature on cost or effort estimation in the context of agile software development.
    \item It reports a systematic literature review study. We rely here on the author's description of the study.
    \item Peer-reviewed papers only.
    \item Written in English only.
\end{itemize}

 After removing 18 duplicates, 144 papers were evaluated in two stages. In the first stage, the first author reviewed the papers based on the title and abstract using the selection criteria. The papers marked as excluded were then reviewed by the second author. When there was a disagreement between the reviewers, the papers were included for a second-stage review. In the second stage, we reviewed a paper for relevance by reading its full text. Data extraction was performed if the papers met the selection criteria after reading the full text. Otherwise, the papers are excluded from further analysis. Ultimately, we included 18 papers in the analysis.

\subsection{Data extraction}
To answer our research questions, we focused on the following facets:
\begin{itemize}
    \item The research questions of the study. Based on the formulation of the research questions, we extracted the main idea of the question. For example, \textit{``What methods have been used to estimate size or effort in ASD''} (RQ1, SLR9) and \textit{``What are the different approaches/techniques used for software cost estimation?''} (RQ1, SLR 10) were coded into types of estimation techniques. \textit{``Which factors most inﬂuence agile effort estimation?''} (RQ3, SLR3) and \textit{``What are the barriers and difficulties in the application of software cost estimation models?''} (RQ2, SLR10) were coded into challenges of estimation.
    \item The justification for conducting the SLR: e.g., claims about lack of coverage, quality (methodological limitations), lack of existing reviews, rapid technological advancements, methodological updates warranting an additional review, different focus than existing reviews, or temporal update.
    \item Quality assessment of the SLRs, based on the five DARE criteria \cite{budgen_what_2020}. The assessment concerns the systematic review process and whether these activities have been performed. Thus, we adopted the three-point scale: yes (1), partly (0.5), and no (0).
    \item The awareness of existing SLRs in the topic. We checked if the study cited previous SLRs as part of their related work or in the discussion section. If the SLRs were part of the primary studies, we would not have considered their awareness. 
\end{itemize}

The unit of analysis in our study is the SLR study itself. For an SLR to be included in the analysis, the authors should perform the actual review process. Thus, studies that performed analysis on the primary studies identified in other SLRs will be excluded, i.e., \cite{britto14}). Moreover, to avoid double-counting \cite{borstler2023double}, if one SLR study was reported in more than one paper, we included the latest publication only. The final version of the dataset is accessible through a publicly available \cite{edison25}.

\section{Results and analysis}
\label{sec:results_analysis}
\subsection{RQ1: Existing SLRs on cost/effort estimation in the agile software development context}
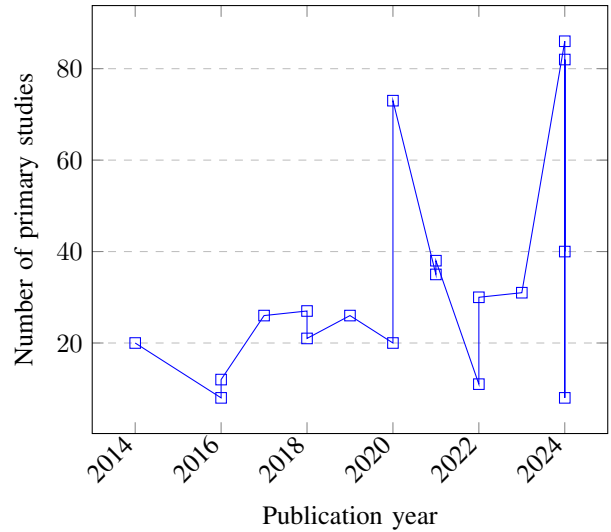
\begin{figure}[h]
    
    \begin{tikzpicture}
        \begin{axis}[
            xticklabel style={rotate=45, anchor=east},
            xlabel={Publication year},
            ylabel={Number of primary studies},
            symbolic x coords ={2014,2015,2016,2017,2018,2019,2020,2021,2022,2023,2024},
            legend pos=north west,
            ymajorgrids=true,
            grid style=dashed,
        ]

        \addplot[
            color=blue,
            mark=square,
        ]
        coordinates {
            (2014,20)(2016,8)(2016,12)(2017,26)(2018,27)(2018,21)(2019,26)(2020,20)(2020,73)(2021,35)(2021,38)(2022,11)(2022,30)(2023,31)(2024,86)(2024,8)(2024,82)(2024,40)
        }; 
        \end{axis}
    \end{tikzpicture}
    \centering
    \caption{Temporal distribution of the SLRs and the number of primary studies included in each of the SLR}
    \label{fig:temporal}
\end{figure}

Based on our search and selection criteria, we identified 18 SLRs on cost/effort estimation in the agile context that were published between 2014 and 2024. Fig. \ref{fig:temporal} shows the distribution of these SLRs over time. Each square represents an SLR. Thus, for a calendar year where two SLRS were published, it will have two squares above the year. The earliest SLR on agile estimation was published in a conference in 2014 (SLR1), and since then, at least one SLR on agile estimation has been published every year except 2015. The peak was reached in 2024 when four SLRS in the area were published. 

In Fig. \ref{fig:temporal}, another dimension is also depicted, which is the number of primary studies included in an SLR. Thus, the placement of a square on the y-axis represents the number of primary studies in that SLR. Only three of the 18 SLRs have more than 57 primary studies, which is an indicator that a research topic has accumulated enough studies that warrant an SLR \cite{wang23}. A majority, i.e., eleven SLRs have between 20 and 40 primary studies. Surprisingly, four papers have an unusually small number of primary studies for an SLR, i.e., 8, 8, 11 and 12. Such a low number of primary studies raises questions about the scope of the SLRS and whether the need and feasibility of the studies were assessed before undertaking the study.

The minimum number of the SLR's primary studies is eight, and the maximum is 86. The mean value is 33, and the median is 26.5, which means the number of reviewed papers is right-skewed. The Spearman rank correlation between the year of publication and the number of reviewed papers is 0.50 (p-value = 0.034), which indicates that the research on cost/effort estimation in agile software development tends to increase moderately over the year. However, we did not check the overlap of primary studies between SLRs.



The number of authors involved in the SLR studies varied between two and eight. The average number of SLR authors is three. One would argue that the reason for involving more persons in an SLR study is to reduce the complexity of reviewing the number of papers returned by the search string from all digital libraries. However, the Spearman rank correlation between the number of papers and the number of authors is 0.194 with a p-value of 0.471. This indicates no strong or significant relationship between the number of authors and the reviewed papers of the SLR.

\subsection{RQ2: Characteristics of the SLRs}
\subsubsection{Aims and objectives}
By coding the research questions of the identified SLRs, we identified eight themes that capture the aims of these SLRs, as shown in Table \ref{tab:SLR_aims} (columns two to nine). Over the years, the main objective of secondary studies has been identifying existing effort/cost estimation techniques (17 reviews). This includes using machine learning models for effort estimation (3 reviews). The second main objective is to assess the performance of effort/cost estimation techniques in terms of their accuracy or effectiveness (12 reviews). In addition, five SLR studies reported the barriers or challenges of effort estimation in agile software development, and four discussed the solutions to address them. 

Effort/cost estimation techniques have been used in different contexts, such as planning activities (SLR1), development activities (SLR1, SLR4, SLR9, SLR16) and mobile apps development (SLR6). Only one study investigates the types of empirical studies on effort estimation, and one study aggregates the benefits of estimation for companies.



\begin{table*}[]
    \centering
    \caption{The aims of the reviews based on the research questions}
    \label{tab:SLR_aims}
\begin{tabular}{p{1.4cm}p{1.3cm}p{1.2cm}p{2cm}p{1.3cm}p{1.3cm}p{1.2cm}p{1.2cm}p{1cm}}
    \\ \hline
    SLR ID & Benefits of estimation for companies & Estimation metrics & Context where effort estimation techniques are applied  & Types of estimation techniques & Assessment of the techniques & Challenges of estimation & Solution to the challenges & Empirical studies \\ \hline
    SLR1 \cite{usman14} &  & \checkmark &  \checkmark & \checkmark & \checkmark & & \\
    SLR2 \cite{bilgaiyan16} &  &  &  & \checkmark & \checkmark & & \\
    SLR3 \cite{schweighofer16} &  &  &  & \checkmark & \checkmark & \checkmark & \\   
    SLR4 \cite{bilgaiyan17} & & \checkmark & \checkmark & \checkmark & \checkmark & & \\ 
    SLR5 \cite{canedo18} & & \checkmark & & \checkmark & \checkmark & & & \checkmark \\ 
    SLR6 \cite{altaleb18} & & \checkmark & \checkmark & \checkmark & \checkmark & & \\
    SLR7 \cite{duran19} & & \checkmark & & \checkmark & & & \\
    SLR8 \cite{arora20} & & & & \checkmark & \checkmark & & \\   
    SLR9 \cite{diego20} & & & \checkmark & \checkmark & \checkmark & & \\ 
    SLR10 \cite{aizaz21} & & & & \checkmark & & \checkmark & & \\ 
    SLR11 \cite{sudarmaningtyas21} & & \checkmark & & \checkmark & \checkmark & & & \\
    SLR12 \cite{alsaadi22} & & & & \checkmark & \checkmark & & & \\
    SLR13 \cite{tandon22} & & & & \checkmark & \checkmark & \checkmark & & \\   
    SLR14 \cite{abusaeed23} & & & & \checkmark & & & & \\ 
    SLR15 \cite{iqbal24} & & & & \checkmark & & \checkmark & \checkmark & \\ 
    SLR16 \cite{moechtar24} & \checkmark & & \checkmark & \checkmark & \checkmark & \checkmark & & \\
    SLR17 \cite{pasuksmit24} & & & & \checkmark &  & & \checkmark &\\
    SLR18 \cite{sinaga24} & & & & & & & \checkmark & \\  \hline
\end{tabular}
\end{table*}

\subsubsection{Years coverage}
\begin{figure}[!h]
    \centering
    \includegraphics[width=0.49\textwidth]{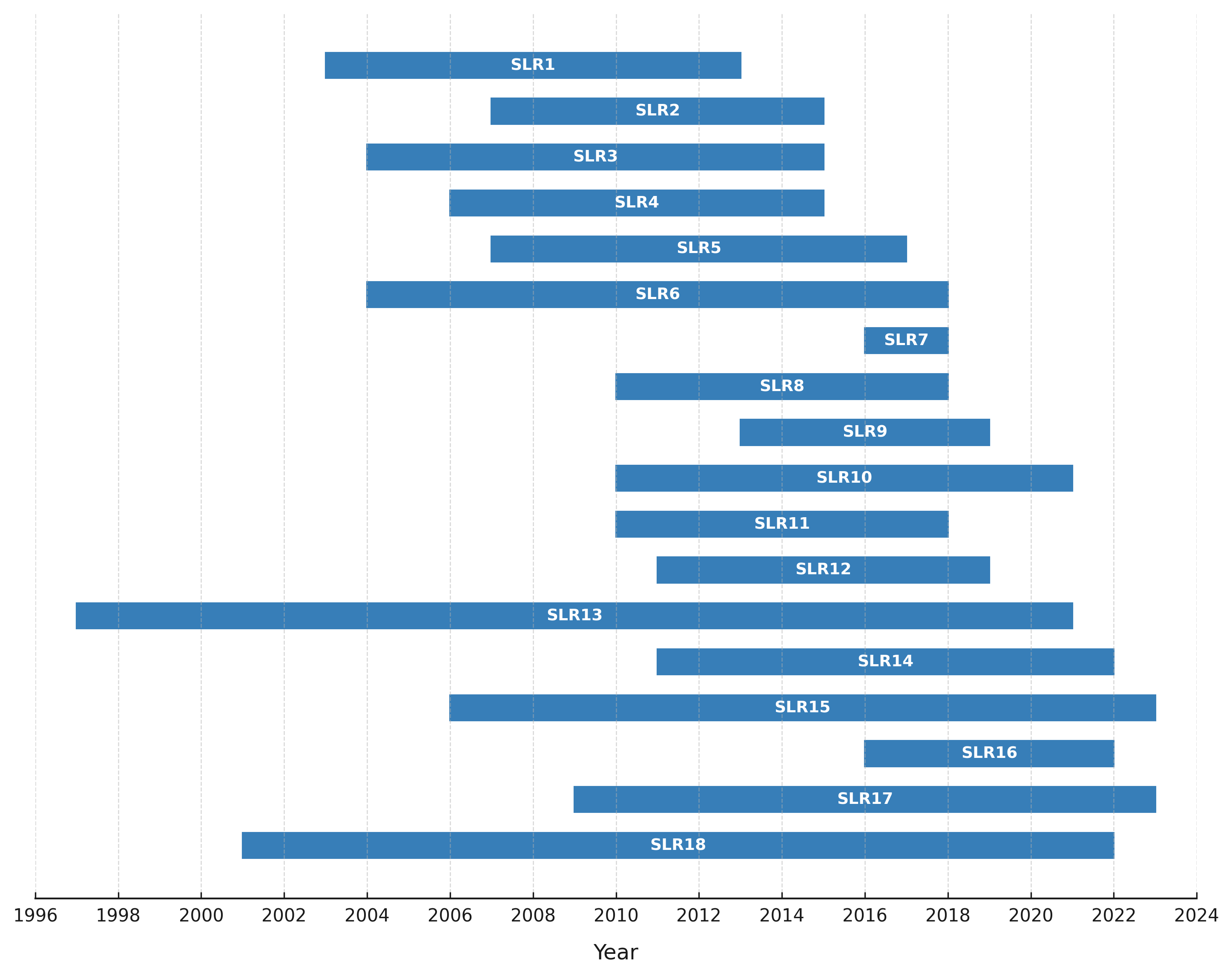}
    \caption{Years coverage of the SLRs}
    \label{fig:years_coverage}
\end{figure}

Fig. \ref{fig:years_coverage} shows the years of coverage of the SLRs on cost/effort estimation in agile software development. The median and mean values of the number of primary studies are 26.5 and 33, with a standard deviation of 23.20. The longest review period covered by the reviewed papers in an SLR is 24 years (SLR13). The SLR was published in the International Journal of Computer Applications. However, only 30 papers were reviewed in the SLR. The SLR showed that the first study on cost/effort estimation in agile software development was published in 1997. 

SLR15 and SRL7 are about user-story-based effort estimation. SLR15 covered 17 years but reviewed 86 papers that were published from 2006 to 2023. Whereas SLR7 covers the shortest time period from 2016 to 2018, and it includes 24 papers.

\subsubsection{Quality score}
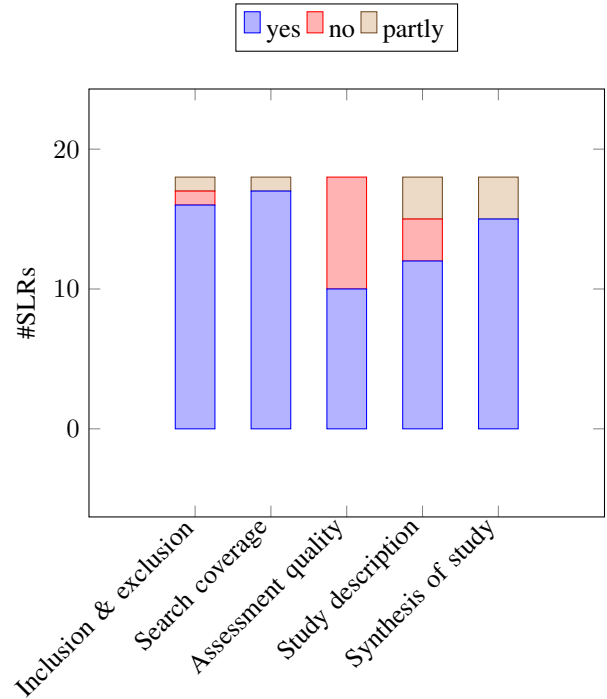
\begin{figure}[!h]
\centering
\begin{tikzpicture}
\begin{axis}[
    ybar stacked,
	bar width=15pt,
    enlargelimits=0.35,
    legend style={at={(0.5,1.2)},
      anchor=north,legend columns=-1},
    ylabel={\#SLRs},
    ymin=0,
    symbolic x coords={Inclusion \& exclusion, Search coverage, Assessment quality, Study description, 
		Synthesis of study,},
    xtick=data,
    x tick label style={rotate=45,anchor=east},
    ]
\addplot+[ybar] plot coordinates {(Inclusion \& exclusion,16) (Search coverage,17)(Assessment quality,10)(Study description,12)(Synthesis of study,15)};
\addplot+[ybar] plot coordinates {(Inclusion \& exclusion,1) (Search coverage,0)(Assessment quality,8)(Study description,3)(Synthesis of study,0)};
\addplot+[ybar] plot coordinates {(Inclusion \& exclusion,1) (Search coverage,1)(Assessment quality,0)(Study description,3)(Synthesis of study,3)};
\legend{\strut yes, \strut no, \strut partly}
\end{axis}
\end{tikzpicture}
 \caption{SLRs' quality score using DARE criteria}
    \label{fig:quality_score}
\end{figure}

The average quality score of the SLRs is 4.1, while the minimum score is two, and the maximum is five. It can be seen in Fig. \ref{fig:quality_score}, among the five criteria to assess the quality of tertiary study\cite{budgen_what_2020}, almost half of the SLRs did not report the quality assessment of their primary studies.

The majority of the SLRs have explicit inclusion/exclusion criteria and use at least four digital libraries such as IEEE Xplore, ACM Digital Library, Scopus, and Google Scholar. Three SLRs did not provide the complete list of the primary studies.

Since the DARE criteria do not question whether an SLR was warranted in the first place or not, the high DARE scores for these papers do not capture the concerns identified in this study about the overlap between the reviewed SLRS.

\subsubsection{Publication ranking}
In our dataset, 56\% of the SLRs (10 reviews) are journal papers while 44\% (8 reviews) are published in conferences. Of the 8 conference papers, only one study (SLR5) was published in a venue with an A rating (Americas Conference on Information Systems) by the CORE Ranking. On the other hand, one SLR was published in an A-rated journal (SLR15), and two in C-rated journals (SLR12, SLR14). The remaining studies (15 studies) were published in non-rated venues. 

Using the Scimago rating classification, four venues of the SLRs were rated Q1 (SLR9, SLR12, SLR14, and SLR17). Two venues were rated Q4 (SLR4 and SLR11). The remaining venues were not rated in Scimago.

\subsection{RQ3: Awareness of the previous SLRs on the topic}
In Table \ref{tab:SLR_citation}, the SLRs are sorted by the year of publication. SLR1 is the first SLR in agile cost and effort estimation while SLR18 is the latest one. The SLRs in the rows cite the SLRs in the columns. For example, SLR1 did not cite any other SLRs, while SLR6 cited SLR1 and SLR4 (only two out of five SLRs that have been published in previous years). 

From the 18 SLRs on cost/effort estimation in the agile context, more than 60\% of them (11 studies) were not aware of previously published SLRs. No citation to previous SLRs was discussed or mentioned in the related work or in the discussion section. For example, three SLRs (SLR7, SLR12 and SLR15) conducted reviews on user story estimation in 2019, 2022, and 2024, respectively. We expect that newer reviews would make an extensive effort to become aware of similar or related work on the topic. Conference papers like SLR7 usually take a shorter period of time to be publicly available than journal papers. However, SLR12 did not cite SLR7, and SLR15 did not cite SLR7 or SLR12.

\begin{table*}[h!]
    \centering
    \caption{Citation of SLRs on cost/effort estimation in the agile context}
    \label{tab:SLR_citation}
\begin{tabular}{p{0.5cm}p{0.4cm}p{0.4cm}p{0.4cm}p{0.4cm}p{0.4cm}p{0.4cm}p{0.4cm}p{0.4cm}p{0.4cm}p{0.5cm}p{0.5cm}p{0.5cm}p{0.5cm}p{0.5cm}p{0.5cm}p{0.5cm}p{0.5cm}p{0.5cm}}
    \\ \hline
     & SLR1 & SLR2 & SLR3 & SLR4 & SLR5 & SLR6 & SLR7 & SLR8 & SLR9 & SLR10 & SLR11 & SLR12 & SLR13 & SLR14 & SLR15 & SLR16 & SLR17 & SLR18 \\ \hline
    SLR1 & -- & & & & & & & & & & & & & & & & &  \\
    SLR2 & & -- & & & & & & & & & & & & & & & &\\
    SLR3 & & & -- & & & & & & & & & & & & & & & \\   
    SLR4 & & & & -- & & & & & & & & & & & & & & \\ 
    SLR5 & & & & & -- & & & & & & & & & & & & & \\ 
    SLR6 & \checkmark & & & \checkmark & & -- & & & & & & & & & & & & \\
    SLR7 & & & & & & & -- & & & & & & & & & & & \\
    SLR8 & & & & & & & & -- & & & & & & & & & &\\   
    SLR9 & \checkmark & & \checkmark & \checkmark & & \checkmark & & & -- & & & & & & & & & \\ 
    SLR10 & & & & & & & & & & -- & & & & & & & & \\ 
    SLR11 & & & & & & & & & & & -- & & & & & & & \\
    SLR12 & \checkmark & & \checkmark & & & & & \checkmark & \checkmark & & & -- & & & & & &\\
    SLR13 & & & & & & & & & & & & & -- & & & & & \\   
    SLR14 & \checkmark & & & & & & & & \checkmark & & & & & -- & & & & \\ 
    SLR15 & \checkmark & & & \checkmark & & & & & \checkmark & & & & \checkmark & & -- & & & \\ 
    SLR16 & & & & & & & & & \checkmark & & & & & & & -- & & \\
    SLR17 & \checkmark & & & & & & & & \checkmark & & & \checkmark & & & & & -- &\\
    SLR18 & & & & & & & & & & & & & & & & & & --\\  \hline
\end{tabular}
\end{table*}


Some of the reviews treated previous SLRs as the primary studies instead, which makes them prone to the threat of double-counting \cite{borstler2023double}. For example, in SLR3, SLR1 is considered one of the primary studies.

\begin{figure}[!h]
    \centering
    \includegraphics[width=0.49\textwidth]{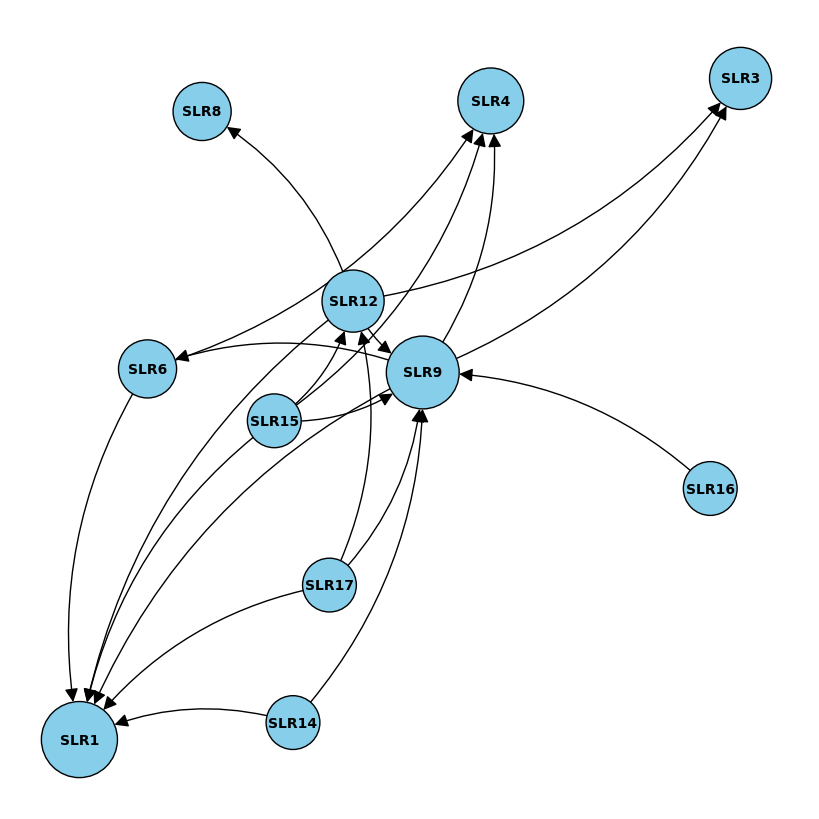}
    \caption{SLRs citation network}
    \label{fig:citation_network}
\end{figure}

Fig. \ref{fig:citation_network} shows the SLRs citation network. The direction of the arrow shows which papers cite which paper. SLR1 and SLR9 are the most cited studies in all reviews, by six and five SLRs, respectively. The authors of SLR9 mentioned that their study was an update of SLR1.  

\subsection{RQ4: Justification for undertaking another SLR on the topic}
The first SLR on the topic (SLR1) was published in 2014 and argued that there were no existing reviews that aggregate the evidence of cost/effort estimation in the agile context. Since then, 17 SLRs on the topic have been published for various reasons. For example, three SLRs claimed that there is a lack of existing reviews in the area of agile estimation for mobile applications (SLR6), estimating the complexity in user stories (SLR7), and machine learning estimation models of Scrum-based projects (SLR8). However, even though SLR11 argued that there was a lack of existing reviews, their first research question is similar to those in SLR1. Yet SLR11 did not cite any previous SLRs, including SLR1.

\begin{displayquote}
\textit{``What kind of method is used for effort estimation in agile?''} -- RQ1, SLR11 \\

\textit{``What techniques have been used for effort or size estimation in agile software development?''} -- RQ1, SLR1
\end{displayquote}

As shown in Table \ref{tab:SLR_agile_estimation}, 44\% of the SLRs (8 studies) did not provide any reasons or justification for why another SLR is needed, even though they were aware of the existing SLRs on the topic. For example, SLR15 discussed several previous SLRs in the area (e.g., SLR1, SLR4, SLR9, and SLR12), but the study did not provide any details on why these SLRs are not sufficient to answer their research questions. Their first research question \textit{(``What are the different methods that are used for effort estimation in ASD?'')} is similar to SLR1's first research question. Similarly, SLR16 did not provide any reason for undertaking another SLR, even though they adopted the search string defined in SLR9 in their study.


\begin{table*}[]
    \centering
    \caption{SLRs on cost/effort estimation in the agile context}
    \label{tab:SLR_agile_estimation}
\begin{tabular}{p{1.4cm}p{1cm}p{5cm}p{2cm}p{1.1cm}p{1.2cm}p{1.2cm}}
    \\ \hline
    SLR ID & Published Year & Claimed justification for another SLR & Cited previous SLRs & Years coverage & Quality score & \# Primary studies \\ \hline
    SLR1 & 2014 & Lack of existing reviews & - & 10 & 5 & 20 \\
    SLR2 & 2016 & No justification  & - & 8 & 4.5 & 8\\
    SLR3 & 2016 &  No justification & - & 11 & 5 & 12 \\   
    SLR4 & 2017 &  No justification & - & 9 & 3 & 26 \\ 
    SLR5 & 2018 & Lack of existing reviews & - & 10 & 2 & 27 \\ 
    SLR6 & 2018 & Lack of existing reviews & SLR1, SLR4 & 14 & 5 & 21 \\
    SLR7 & 2019 & Lack of existing reviews \& Different focus than existing reviews & - & 2 & 5 & 26 \\
    SLR8 & 2020 & Lack of existing reviews & - & 8 & 5 & 20\\   
    SLR9 & 2020 & Temporal update & SLR1, SLR3, SLR4, SLR6,  & 6 & 4 & 73 \\ 
    SLR10 & 2021 & No justification & - & 11 & 5 & 35\\ 
    SLR11 & 2021 & Lack of existing reviews & - & 8 & 4 & 38 \\
    SLR12 & 2022 & Rapid technology advancement \& Different focus than existing reviews & SLR1, SLR3, SLR8, SLR9, & 8 & 2.5 & 11\\
    SLR13 & 2022 & No justification & - & 24 & 4.5 & 30\\   
    SLR14 & 2023 & Temporal update & SLR1, SLR9 & 11 & 5 & 31 \\ 
    SLR15 & 2024 & No justification & SLR1, SLR4, SLR9, SLR12 & 17 & 3 & 86 \\ 
    SLR16 & 2024 & No justification & SLR9 & 6 & 5 & 8 \\
    SLR17 & 2024 & Different focus than existing reviews & SLR1, SLR9, SLR12 & 14 & 3 & 82\\
    SLR18 & 2024 & No justification & - & 21 & 3.5 & 40 \\  \hline
\end{tabular}
\end{table*}

\section{Discussion}
\label{sec:discussion}
\subsection{Redundancy and lack of clear justification of additional systematic reviews}
Our findings underscore several critical issues prevalent in systematic literature reviews (SLRs) within the context of agile software estimation. Firstly, there is a noticeable lack of awareness among authors regarding existing literature reviews, with over 60\% of the analyzed studies failing to cite or discuss prior relevant SLRs. This suggests a significant gap in the initial literature assessment process, indicating that authors either overlook or inadequately assess existing knowledge before initiating new reviews.

Moreover, justification for conducting new SLRs was frequently absent or insufficiently articulated. Half of the reviewed SLRs did not explicitly justify the necessity of their study, while those providing justifications often cited superficial or redundant reasons such as a ``lack of existing reviews,'' despite clear evidence to the contrary. This pattern reveals a critical need for a stronger emphasis on explicit justification and validation when proposing new secondary studies.

This situation is not unique for literature reviews on cost and effort estimation in Agile. Similar observations have been made by others in other topics such as regression testing \cite{ali2019search}, software analytics \cite{laiq2024software,laiq2025} and DevOps \cite{hrusto_enhancing_2024}, where a large number of secondary studies have been published. Cost and effort estimation in the context of Agile software development presents a fairly niche topic, which makes it an interesting case to study, with 18 published reviews on an apparently narrow topic. 

Like any research, it is important for secondary studies to capitalize on existing research. However, the effort-intensive and time-consuming nature of SLRs \cite{zhang2013systematic} makes it even more important. 

\begin{figure}
    \centering
    \includegraphics[width=0.80\linewidth]{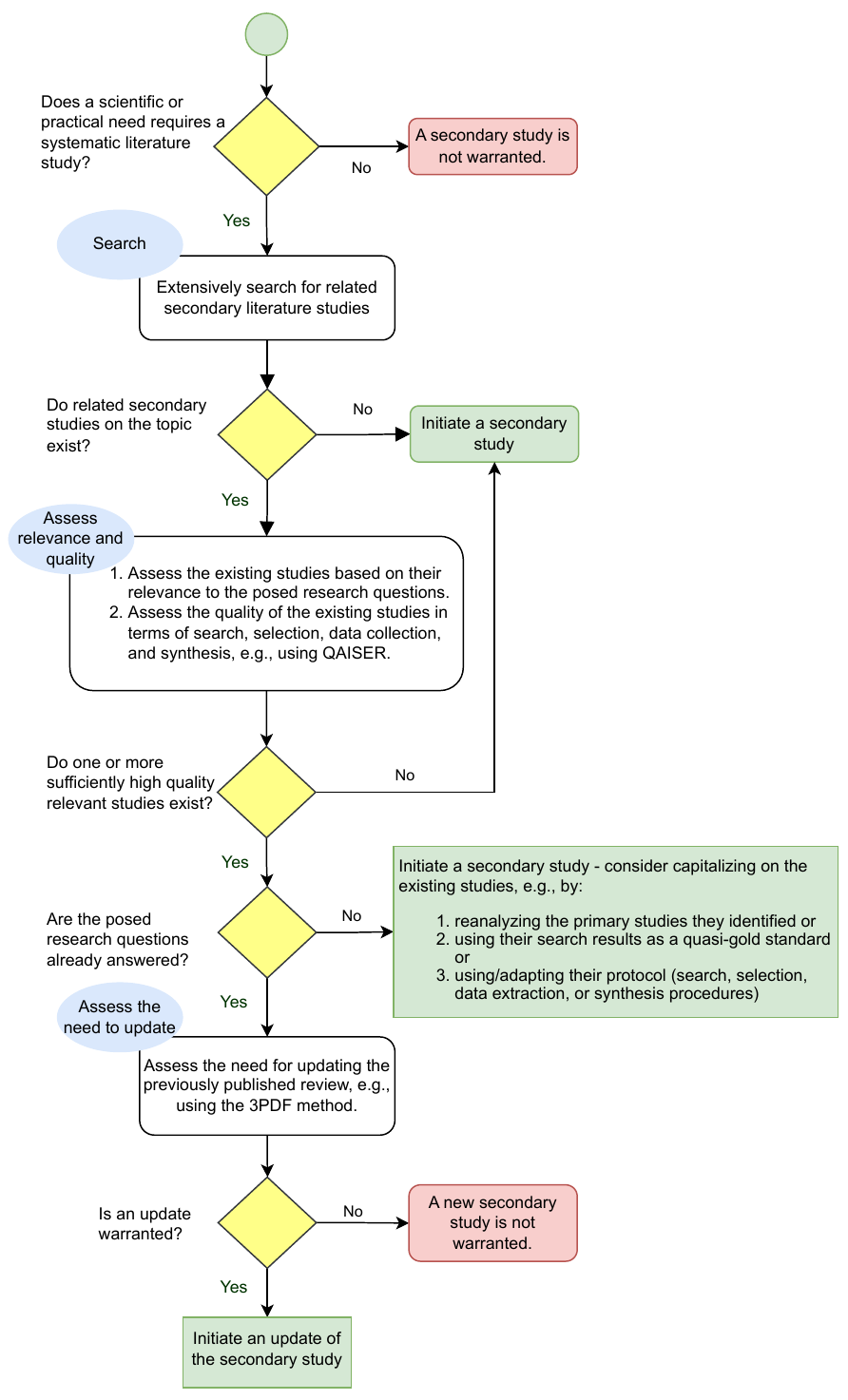}
    \caption{Work flow to systematically identify and utilize existing systematic literature reviews.}
    \label{fig:workflow}
\end{figure}

\subsection{Guidelines for searching and analyzing related work}
The guidelines for conducting secondary studies already emphasize the importance of establishing the need for undertaking a literature review \cite{kitchenham2015evidence}. Furthermore, a recent quality appraisal instrument for systematic literature reviews enables reviewers to assess whether a study has sufficiently justified the need \cite{Usman2023}. In Figure \ref{fig:workflow}, we further strengthen the guidelines for researchers by presenting a workflow for systematically approaching and dealing with existing secondary studies.

In software engineering, there is no prospective register\footnote{\href{Prospero}{https://www.crd.york.ac.uk/prospero/}} for announcing and describing a literature review that a research group is about to undertake. So, there will always be some unintended duplication of effort. However, our proposed workflow (see Figure \ref{fig:workflow}) intends to reduce the likelihood of overlooking or not capitalizing on existing secondary studies.

The three main activities in this workflow are 1) search for existing reviews, 2) assessment of existing reviews, and 3) assessment of the need for an update. 
\begin{enumerate}
\item \textbf{Search:} For search, we propose a general search string that is composed of two blocks of keywords joined by a Boolean `and'.

\textbf{Block 1:} keywords to identify systematic literature studies including: ``systematic review'' OR ``research review'' OR ``research synthesis'' OR ``research integration'' OR ``systematic overview'' OR ``systematic research synthesis'' OR ``integrative research review'' OR ``integrative review'' OR ``systematic literature review'' OR ``literature review'' OR ``scoping review'' OR ``systematic map'' OR ``systematic mapping'' OR ``mapping study'' OR ``tertiary study'' OR ``tertiary review'') \footnote{Although we are looking for existing systematic literature reviews, we recommend that the researchers make a more inclusive search for non-systematic literature reviews, mapping studies, and any tertiary studies.}

\textbf{\textit{AND}}

\textbf{Block 2:} topic-specific keywords, e.g., in this study, we used effort estimation and agile-related keywords.

\item \textbf{Assess relevance and quality:} For the assessment of existing reviews, we recommend the use of Quality Assessment Instrument for Software Engineering systematic literature Reviews (QAISER) \cite{Usman2023}. It is a specialized tool that adapts the healthcare-focused AMSTAR 2 assessment instrument, tailoring it specifically for software engineering contexts. QAISER incorporates 15 evaluation criteria that assess key elements of SLR quality, such as the clarity and relevance of research questions, the comprehensiveness of literature searches, the appropriateness of methods for study selection, data extraction, and synthesis procedures.

A structured evaluation process within QAISER allows reviewers to systematically identify strengths and weaknesses in the SLR's need, methodology, and findings. 

The development of QAISER included validation through input from four external software engineering experts (including leading guideline authors like Prof. Barbara Kitchenham and Prof. Kai Petersen). Its utility has been demonstrated through practical applications where the tool was successfully employed to appraise six systematic literature studies. 

\item \textbf{Assess the need to update:} In the third activity in the workflow, when deciding whether a previous review should be updated, we suggest the use of the guidelines by Mendes et al. \cite{mendes_when_2020}.

\end{enumerate}

\subsection{Threats to validity}
Threats to validity related to the results from this study are discussed below.

\subsubsection{Search comprehensiveness}
We only constrained our study to SLR and did not look for mapping studies. Our search strings also yielded some mapping studies, and those were excluded. This means that we may have overlooked relevant papers. However, we think having already identified 18 relevant SLRs is not a major threat to the findings of this study.

\subsubsection{Researchers bias} 
To reduce the researchers' bias, two reviewers were involved in the selection of the SLRs and in judging the claimed justification reported in the papers. One reviewer performed the data extraction using an extraction form. Both reviewers then discussed and evaluated a pilot with three papers. Any ambiguities and confusion about the extraction process were resolved.

\subsubsection{Generalization}
In this study, we only focus on SLRs on one topic, which is cost/effort estimation in agile software development. Thus, the problem may not be generally true in other secondary studies in other topics within software engineering domain.



\section{Conclusion}
\label{sec:conclusion}

This study critically examined the landscape of systematic literature reviews (SLRs) on effort and cost estimation in agile software development. Our analysis of 18 SLRs revealed substantial redundancy, insufficient awareness of prior reviews, and a lack of clear justification for conducting new secondary studies. These findings highlight systemic issues in how SLRs are initiated and positioned within the empirical software engineering community.

To mitigate these challenges, we propose a structured workflow to approach related work comprising three key activities: a comprehensive search for related reviews, rigorous quality appraisal using tools such as QAISER, and a principled evaluation of the need for an updated SLR. This workflow aims to support researchers in making evidence-informed decisions about initiating secondary studies and to enhance the transparency and utility of their contributions.

Our findings contribute to the broader discourse on improving research practices in software engineering by emphasizing the importance of methodological justification, reuse of prior evidence, and editorial scrutiny. We call upon the community—researchers, reviewers, and venue organizers alike—to adopt and reinforce practices that prevent redundant research and promote cumulative, high-quality evidence synthesis.

Future work can be done by broadening our sample by collecting and analysis the systematic mapping studies or any other types of secondary studies in cost/effort estimation in agile software development. Thus, with a more comprehensive search, the findings will complement our study. Another venue for future work is to replicate our study in other topics within software engineering. By contrasting and comparing the secondary studies in various topics, we can improve the generalizability of the findings. 

Future work could also further investigate the list of primary studies identified by the SLRs in our sample. We will explore the overlap between the primary studies of the SLRs that focus on the same research questions. Such a study could confirm or refute our findings.

\section*{Acknowledgements}
This work was supported by ELLIIT, the Swedish government-funded Strategic Research Area within IT and Mobile Communications, and a research grant for the GIST project (reference number 20220235) from the Swedish Knowledge Foundation.

\bibliographystyle{IEEEtran}
\bibliography{references}

\end{document}